# Beaming Elastic and SERS Emission from Bent-Plasmonic Nanowire on a Mirror Cavity

Sunny Tiwari,*,† Adarsh B. Vasista,‡ Diptabrata Paul,† Shailendra K. Chaubey,† and G.V. Pavan Kumar*,†

†Department of Physics, Indian Institute of Science Education and Research, Pune-411008, India

‡Department of Physics and Astronomy, University of Exeter EX44QL, United Kingdom

*E-mail: sunny.tiwari@students.iiserpune.ac.in; pavan@iiserpune.ac.in

## Abstract

We report on the experimental observation of beaming elastic and surface enhanced Raman scattering (SERS) emission from a bent-nanowire on a mirror (B-NWoM) cavity. The system was probed with polarization resolved Fourier plane and energy-momentum imaging to study the spectral and angular signature of the emission wavevectors. The out-coupled elastically scattered light from the kink occupies anarrow angular spread. We used a self-assembled monolayer of molecules with a well-defined molecular orientation to utilize the out-of-plane electric field in the cavity for enhancing Raman emission from the molecules and in achieving beaming SERS emission. Calculated directionality for elastic scattering and SERS emission were found to be 16.2 and 12.5 dB respectively. The experimental data were corroborated with three-dimensional numerical finite element and finite difference time domain based numerical simulations. The results presented here may find relevance in understanding coupling of emitters with elongated plasmonic cavities and in designing on-chip optical antennas.

Directional optical antennas are at the heart of nano-photonics as they influence and provide control on the properties of light for efficient detection and on-chip coupling.[1–3] Thus, there is a continuous endeavor to design structures which can scatter light directionally with a narrow angular spread. To this end, various metallic structures supporting surface plasmons have been utilized to confine and scatter light efficiently at a subwavelength scale.[4–7]

One-dimensional nano-structures such as plasmonic nanowires[5,8] and bent-nanowires[9] have shown to influence the directionality of the scattered light[10,11] and have been used utilized in remote detection of molecules,[12] strong coupling physics,[13] and as a subwavelength channel of molecular emission.[11,14] Plasmonic cavities that are formed by coupling two plasmonic structures enhance light-matter interaction by confining the optical field to volumes less than 1 nm$^3$.[15–19] Thus, plasmonic nano-cavities are utilized to probe single molecule-surface enhanced Raman scattering (SERS),[20,21] strong coupling physics,[22] and in designing optical sensors.[23] Of late, mirror based plasmonic cavities have gained special attention because of the ease in preparation and the ultra-small mode volume supported by them.[24,25] Metallic substrates direct maximum emission towards the collection objective which increases the sensitivity of enhanced spectroscopy techniques, by increasing signal to noise ratios.[26,27] In the past, structures with the optical properties of a directional antenna have mostly been used on substrates with a high refractive index which leads to leakage of majority of emission into the substrates.[5,28,29] The collected emission through the substrate has a broad angular distribution in the azimuthal angles around substrate-air critical angle. In addition, leakage of the optical signal into the substrate reduces the directionality of the antenna because of the presence of the leaky modes.[29]

Directional antennas have also been used to enhance, detect and direct molecular emission such as fluorescence and SERS signals in a direct or a remote excitation configuration.[9,12,30] But, there was a minimal control over placement and orientation of the molecules on these structure which reduced the directionality of the antenna by increasing the angular spread of emission. Moreover, since most of these geometries are individual structures, the field enhancement supported by them are minimal because of the absence of hotspots. Thus, there is a demand to design geometries which can scatter elastic and inelastic light with high directionality and narrow angular distribution, without compromising the field enhancement.

Motivated by this, we design and utilize a bent-plasmonic nanowire placed on a gold mirror for directing elastically scattered light and SERS emission. We show beaming of the elastically scattered light and the SERS from the bent-nanowire to a very narrow range of wavevectors with the excitation and the collection points separated by few microns. We used a monolayer of

biphenyl-4-thiol molecules with an out-of-plane orientation self-assembled on the mirror to utilize maximum field enhancement in the cavity for SERS enhancement and in achieving beaming SERS. We utilized Fourier plane and energy-momentum imaging[31–33] to probe the wavevectors of out-coupled elastically scattered light and SERS from the geometry. In addition, we performed three-dimensional numerical simulation to understand and corroborate the experimental results.

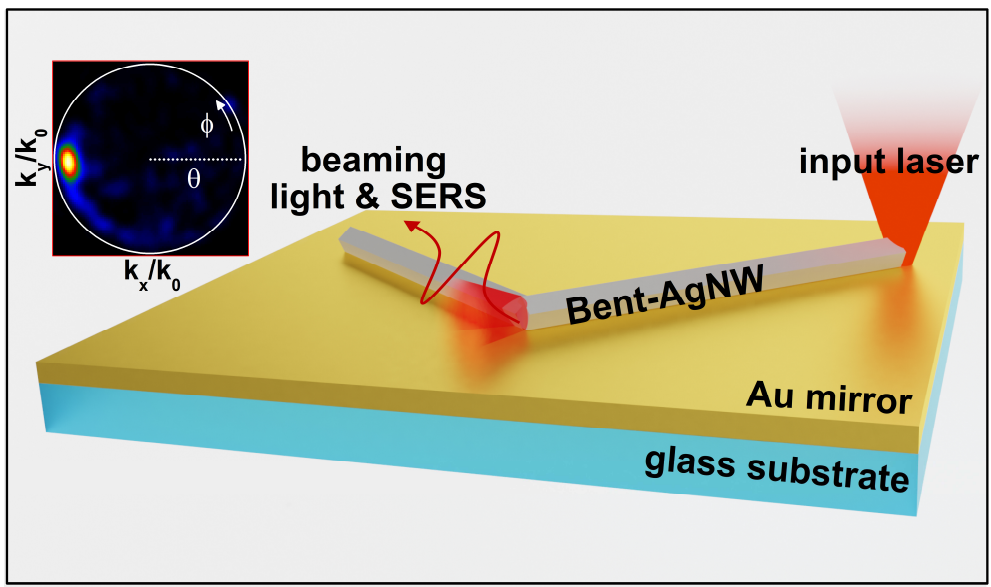

**Figure 1:** Schematic of the experimental configuration. A bent silver nanowire was placed on a gold mirror and one end of the nanowire was excited with a focused laser. The out-coupled light from the kink of the nanowire was spatially filtered and was projected onto the EMCCD for Fourier plane imaging. For SERS experiments, a monolayer of biphenyl-4-thiol molecules was sandwiched between the wire and the mirror. The light out-coupled from the kink was projected onto the EMCCD and spectrometer for Fourier plane and energy-momentum imaging after rejecting the laser line.

A schematic of the experimental configuration is shown in figure 1. A single crystalline, chemically synthesized[34] bent silver nanowire (AgNW) was placed on a gold mirror making an extended bent-nanowire on mirror cavity (B-NWoM cavity). A description of the bending of the nanowire using ultrasonication can be found elsewhere.[35] One end of the nanowire was excited with a tightly focused 633 nm laser beam using a high numerical aperture objective lens which excites nanowire surface plasmon polaritons (SPPs) along with the junction plasmon modes between the nanowire and the mirror. The plasmon polaritons propagating along the nanowire out-couples as free-space photons at the kink. The scattered light from the kink was spatially filtered and was projected onto the EMCCD for Fourier plane imaging. For SERS studies, a monolayer of biphenyl-4-thiol (BPT) molecules was assembled on the gold mirror, using self-assembled monolayer technique, following that bent nanowires were dropcasted. The system was then excited on one end with a focused 633 nm laser beam. Out-coupled light from the kink was spatially filtered and was projected onto the EMCCD/spectrometer for Fourier plane and energy-momentum imaging and spectroscopy after rejecting the laser line. (See supplementary information S1 for sample preparation, and S2 for detailed experimental setup.)

Figure 2(a) shows bright field image of a bent-NW with an inter-arm angle α=106°, placed on a 160 nm thick gold mirror. One end of the nanowire was excited with a 633 nm laser with polarization along the axis of the nanowire using a high numerical aperture objective lens (see figure 2(a)(ii)). Nanowire SPPs scatter out as free space photons at the kink. We collected the emission only from the kink using a spatial filter and projected the light to the EMCCD to perform Fourier plane imaging which maps the angular distribution in terms of radial and azimuthal angles. The Fourier plane image shown in figure 2(b) indicates that the wavevector spread of out-coupled elastic scattered photons are narrow in terms of radial and azimuthal angles. This implies that scattered light is beaming in the form of a small lobe and it is directed towards higher k-values. To quantify the emission, we define directionality (Dir), using the ratio of forward and backward intensity of emission in the Fourier plane,[6] as

$$\text{Dir} = 10\log_{10} \frac{\iint_{(\theta_m-\delta_1,\phi_m-\delta_2)}^{(\theta_m+\delta_1,\phi_m+\delta_2)} I(\theta,\phi)\sin(\theta)d\theta\,d\phi}{\iint_{(\theta_m-\delta_1,\phi_m-\pi-\delta_2)}^{(\theta_m+\delta_1,\phi_m-\pi+\delta_2)} I(\theta,\phi)\sin(\theta)d\theta\,d\phi} \qquad (1)$$

where $\theta_m$ and $\phi_m$ are the radial and azimuthal angles with maximum emission. I ($\theta$, $\phi$) is the intensity in the Fourier plane image. For the white dotted region in the Fourier plane image (see figure 2 (b)), with $\delta_1$=7.5° and $\delta_2$=10° the calculated directionality of the emission is (16.2 ± 0.1) dB, which is an excellent number for a structure prepared using bottom-up approach.

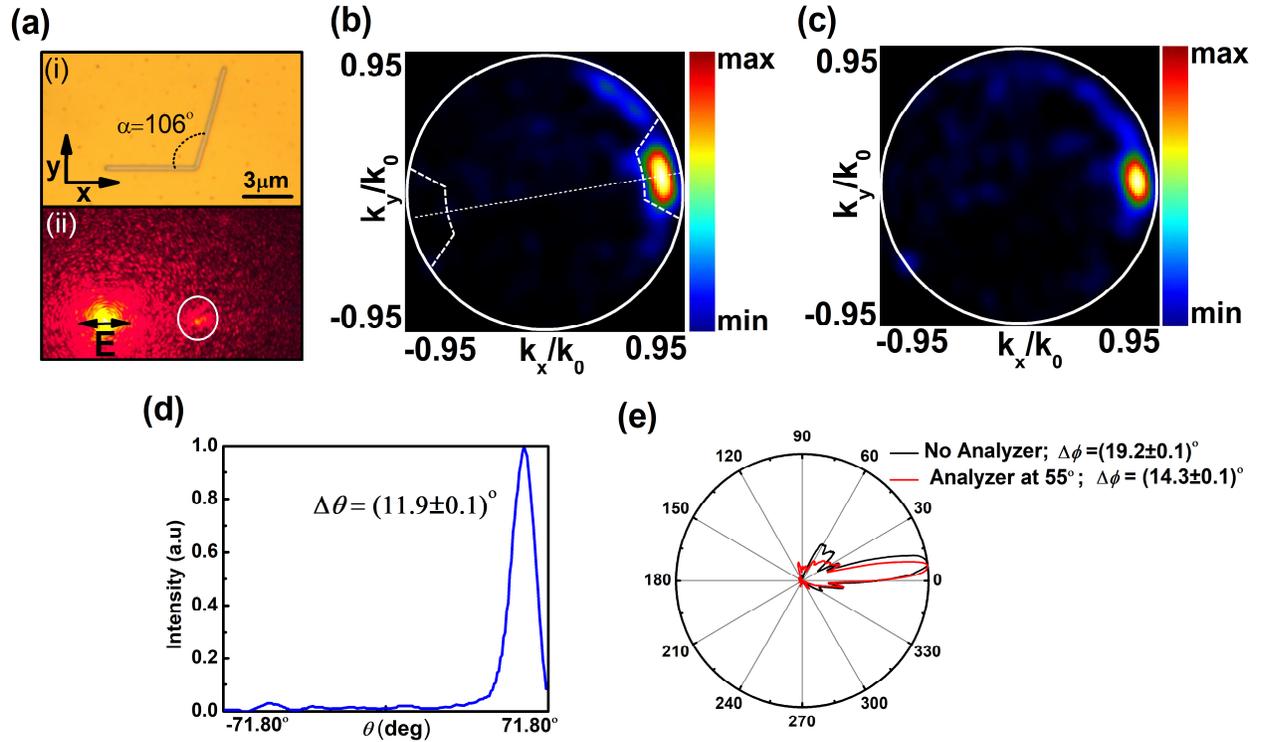

**Figure 2:** Directional elastic emission from B-NWoM geometry. (a) (i) Bright field optical image of a bent-nanowire with an inter-arm angle α=106° placed on a 160 nm thick gold mirror. (ii) Corresponding image of the same nanowire excited at one end with a 633 nm laser. (b) Fourier plane image of elastic scattering light from the kink of the nanowire (shown as white circle in (a)(ii)). (c) Polarization resolved Fourier plane image when the emission is analyzed at an angle of 55° with respect to nanowire axis. (d) Intensity cross-cut along the white dotted line in Fourier plane image showing the confinement of light with a FWHM, Δθ, of 11.9°. (e) The intensity profile of azimuthal angles (ϕ) for θ corresponding to maximum intensity in the Fourier plane images (c) and (d). The FWHM are only 19.2° and 14.3° for unanalyzed and analyzed emission respectively.

In addition to high directionality, an antenna should provide narrow angular distribution along the azimuthal angles. Various structures acting as directional optical antennas have shown high directionality but the spreading in the azimuthal angles of out-coupled emission is very large.[5,9,30] The unwanted signatures of leaky modes are also present in the far-field radiation patterns, which reduces the directionality and angular confinement of emission, for example, bent-nanowires placed on glass substrate shows considerably large angular spread along the azimuthal angles (See supplementary information S3 for experiments performed on bent-nanowire on glass substrate). Furthermore, collecting the emission through the air side reduces the collection efficiency, as majority of the emission leaks into the substrate having higher refractive index.[4] The gold substrate used in our experiments acts as a mirror thus minimizes the photon loss due to leakage while decreasing the angular spread of emission.

Intensity cross-cut (figure 2(d)) along the white dotted line in figure 2(b) shows that the emission is confined to a narrow range of radial angles with a full width at half maxima (FWHM) ($\Delta\theta$) of only $(11.9 \pm 0.1)°$. In addition, the intensity profile of azimuthal angles ($\phi$) for $\theta$ corresponding to maximum intensity in the Fourier plane image shows that the majority of emission is going towards higher wavevectors with a very narrow range of azimuthal angles with a FWHM ($\Delta\phi$) of only $(19.2 \pm 0.1)°$(see figure 2(e)). Although the propagation of nanowire plasmon polaritons is along the nanowire axis which is along $k_y/k_0=0$ the presence of the kink shifts the out-coupled emission towards higher $+k_y/k_0$.

For straight silver nanowires, the emission wavevectors of the out-coupled SPPs from the distal end show interesting polarization signatures.[28] When the out-coupled light from the distal end is analyzed, for a particular polarization the emission is relatively narrow (see supplementary information S4). We show how at a certain angle of analyzer, the angular spread of the emission from B-NWoM geometry can be further reduced. For this, we performed polarization resolved Fourier plane imaging on elastic scattering light from the bent-nanowire. The FWHM of azimuthal spreading in the emission wavevectors reduces from 19.2° to 14.3° at an analyzer angle of 55° with respect to the nanowire axis. The variation of emission wavevectors in Fourier plane images and change in $\Delta\phi$ with respect to analyzer angle is shown in supplementary information S5. See supplementary information S6 for the value of analyzer angle for obtaining minimum $\Delta\phi$ spreading with a change in the inter-arm angles α. For bent-nanowire, with obtuse inter-arm angles, the value of analyzer for obtaining minimum $\Delta\phi$ spreading is approximately half of the inter-arm angle of the nanowire. Whereas for acute inter-arm angles, the value of analyzer angle for minimum $\Delta\phi$ spreading changes peculiarly which shows that the far-field radiation pattern and polarization dependence of emission depends on the exact geometry of the kink of the bent-nanowire.

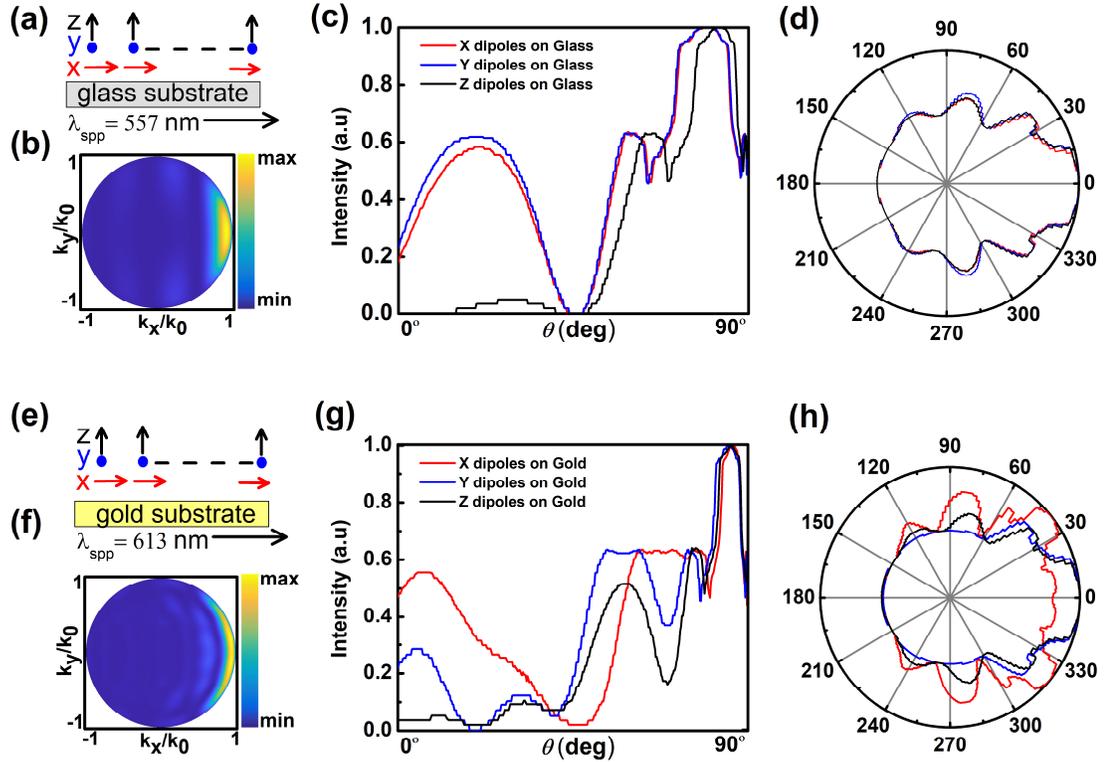

**Figure 3:** Effect of substrate on the far-field radiation pattern of the emission from a chain of dipoles. (a) A chain of 10 dipoles (x, y, or z oriented) placed at a distance of $\lambda/200$ from glass substrate. The phase retardation between the dipoles are set according to the plasmon propagation wavelength of 557 nm. (b) Calculated Fourier plane image for a series of z oriented dipoles on glass substrate. (c) Intensity cross-cuts along the $k_y/k_0 = 0$ in the calculated Fourier plane images for a series of x, y, and z oriented dipoles. (d) Intensity profiles of azimuthal angles ($\phi$) for $\theta$ corresponding to maximum intensity in the Fourier plane images. (e) Chain of 10 dipoles (x, y, or z oriented) placed at a distance of $\lambda/200$ from gold substrate. The phase retardation between the dipoles are set according to the plasmon propagation wavelength of 613 nm. (f) Calculated Fourier plane image for a series of z oriented dipoles on gold substrate. (g) Intensity cross-cuts along the $k_y/k_0 = 0$ in the calculated Fourier plane images for a series of x, y, and z dipoles. (h) Intensity profiles of azimuthal angles ($\phi$) for $\theta$ corresponding to maximum intensity in the Fourier plane images. The complete length of the chain is 1 µm and the dipoles are oscillating at 633 nm.

To understand the effect of substrate on the far-field wavevector distribution of emission from B-NWoM, we studied far-field radiation pattern from a chain of dipoles, to model the waveguiding of light, placed on a glass[5] and a gold substrate, at a distance of $\lambda/200$ from the substrate and oscillating at 633 nm in x, y, or z directions. Although, the exact emission pattern depends on the geometry of the waveguide, yet the effect of substrate on the waveguiding of light using these simulations can be understood. A more involved modelling using a silver nanowire is discussed in the following section. The length of the chain was 1 µm and the distance between consecutive dipoles was 100 nm. The phase retardation between the dipoles was set according to the plasmon propagation wavelength calculated using COMSOL Multiphysics. The details of the finite element method (FEM) based simulation in COMSOL Multiphysics is discussed in the supplementary information S7. For calculating the far-field radiation pattern, the near-field electric field, calculated using FEM based simulations, was transformed to the far-field using reciprocity arguments.[36] The refractive indices of the material were taken from ref.[37] Figure 3(a) and (e) show the geometry of the system used in the simulations. The calculated Fourier plane image for a chain of dipoles, oriented along z axis, placed on a glass substrate and oscillating at 633 nm is shown in figure 3(b). The emission is confined to higher $+k_x/k_0$ values which is along the direction of wave propagation. Intensity cross-cuts along the $k_y/k_0 = 0$ value in Fourier plane images for a chain of x, y, and z oriented dipoles show that the emission is more confined in terms of radial angles when the orientation of dipoles is along z (see figure 3 (c)). The calculated Fourier plane images for chain of dipoles oriented along x and y axis is shown in supplementary information S8. The in-plane oriented dipoles give relatively broader emission. Comparison between the intensity profile of azimuthal angles ($\phi$) for $\theta$ corresponding to maximum intensity in the Fourier plane images for x, y, and z oriented dipoles shows that the spreading is approximately same for all the orientations (see figure 3 (d)).

The above mentioned results suggest that for a waveguide placed on a substrate, the emission will be more confined in the Fourier plane, for an intense z field, which can be achieved by using a gold substrate. For this, we simulated the far-field radiation pattern from a chain of x, y, and z oriented dipoles placed on a gold substrate at a distance of $\lambda/200$ from the substrate. Calculated Fourier plane image for a chain of z oriented dipoles is shown in figure 3(f) and Fourier plane images for x and y oriented dipoles are shown in supplementary information S8. Even for a gold substrate, the emission is more confined in radial and azimuthal angles for z oriented dipoles as compared to in-plane x or y oriented dipoles (see figure 3(g) and 3(h)). The results suggest the importance of gold substrate and intense z field in providing better directionality for waveguides.

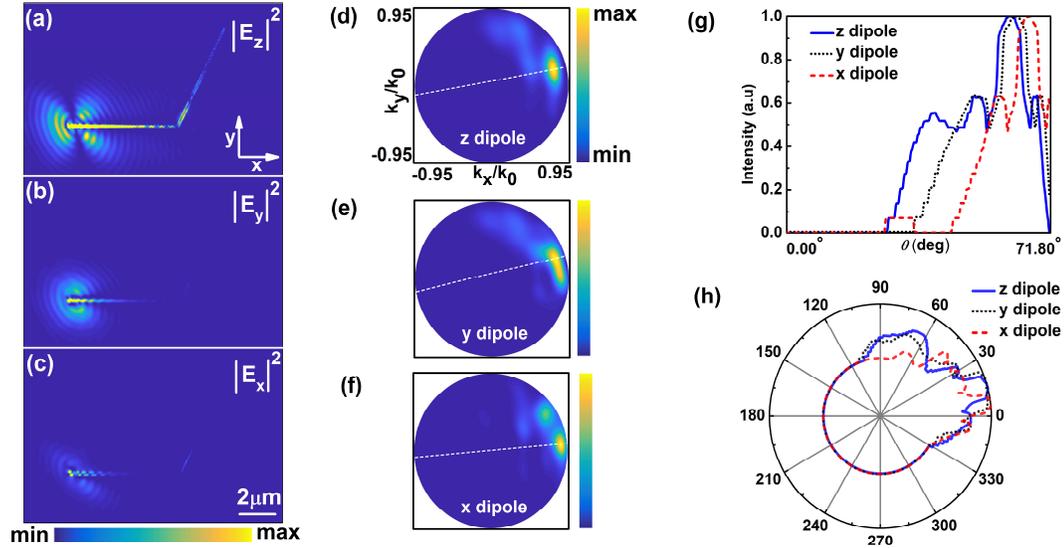

**Figure 4:** Calculated component wise near-field electric field magnitude in the B-NWoM cavity and Fourier plane images of emission from the kink upon excitation of one end of nanowire with different orientation of molecules. (a-c) Calculated component wise near-field electric field inside the cavity for magnitude of $E_z$, $E_y$, and $E_x$ respectively, upon excitation of nanowire end with 633 nm Gaussian excitation with polarization along the nanowire length. (d-f) Calculated Fourier plane images of the emission from kink when nanowire end is excited with z, y, and x dipoles respectively by individually placing a dipole at a wavelength of 703 nm in the B-NWoM cavity. (g) Intensity cross-cuts along the white dotted line in Fourier plane images (d-f). Only positive angular values are shown. (h) The intensity profile of azimuthal angles ($\phi$) for $\theta$ corresponding to maximum intensity in the Fourier plane images (d-f).

Using three-dimensional electrodynamics simulations, we studied how B-NWoM cavity can be used to enhance and direct secondary emission from molecules. For calculating near-field electric field, we performed three-dimensional finite difference time domain (FDTD) calculations in Lumerical software. The details of the geometry and other parameters used in the simulations are given in supplementary information S9. Figure 4a-c show the component wise magnitude of near-field electric field ($E_z$, $E_y$, and $E_x$ respectively) in the B-NWoM cavity when one end of the nanowire was excited with a focused Gaussian beam of wavelength 633 nm with polarization along the length of nanowire. Though the cavity is excited with a laser having in-plane polarization along the nanowire, the magnitude of out-of-plane, $E_z$ field in the cavity is much stronger than in-plane $E_x$ and $E_y$ fields. The results show that for utilizing the maximum field present in the cavity for enhanced scattering, the orientation of the molecule should be out-of-plane or in the z direction.

Next, we studied for which orientation of the dipole, the out-coupled emission from the kink is more directional and confined. For this, we calculated Fourier plane images of emission from the kink. We placed dipoles with x, y, and z orientations in the B-NWoM cavity, near the end of the nanowire and projected the near-field electric field of the kink region to the far-field. Figure 4(d-f) show the calculated Fourier plane images from the kink region of the B-NWoM cavity, when an individual z, y, or x dipole oscillating at a wavelength of 703 nm is placed in the cavity respectively. The wavelength of dipole oscillation is chosen to be 703 nm as the BPT molecules used in the SERS experiments has a prominent Raman mode at 703 nm. Quantitatively, the intensity cross-cuts along the white dotted lines in the Fourier plane images (figure 4(d-f) shows that the emission is more confined in the radial direction for x and z dipoles as compared to the y dipole (see figure 4(g)). The intensity profile of azimuthal angles (figure 4(h)) ($\phi$) for $\theta$ corresponding to maximum intensity in the Fourier plane images (d-f) show that the azimuthal spreading for the z and x dipoles is relatively less as compared to y dipole. This makes the z oriented dipoles to be extremely beneficial for utilizing the B-NWoM cavity for enhancement and high directionality as the $E_z$ field is more intense in the cavity and the emission is narrower when the dipole orientation in the cavity is along z direction.

To utilize the B-NWoM cavity for enhancing and directing SERS emission from molecules, we used a monolayer of vertically oriented self-assembled monolayer of BPT molecules on gold mirror. Figure 5 shows a bent-nanowire with an inter-arm angle of 133° dropcasted on a gold mirror over which there is a self-assembled monolayer of vertically oriented BPT molecules. One end of the nanowire was excited using a high numerical aperture objective lens with a 633 nm laser with polarization along the nanowire (figure 5(a)(ii). Due to the presence of cavity between the nanowire and mirror the molecules undergo SERS emission which gets couples to the nanowire plasmons. These plasmons get out-coupled as free-space photons from the kink. In addition, the SPPs propagating along the nanowire also get out-coupled from the kink of the nanowire, exciting the molecules present at the kink. The out-coupled emission from the kink was spatially filtered and projected to the spectrometer after rejecting the elastic scattered light for spectroscopy and energy-momentum imaging and to EMCCD for Fourier plane imaging. The spectrum (figure 5(b)) collected from the kink of the nanowire shows the SERS signatures of BPT molecules riding over a broad background. To study the directionality of the wavevector of remotely collected SERS spectrum from the BPT molecules, we performed Fourier plane imaging on SERS emission by spatially filtering the kink region and projecting it to EMCCD. The Fourier plane image (figure 5(c)) shows that the SERS emission is beaming towards higher k-values in a small lobe having very narrow angular spreading. The emission is slightly shifted towards higher $+k_y/k_0$ values and is not along the axis of nanowire because of the presence of kink.

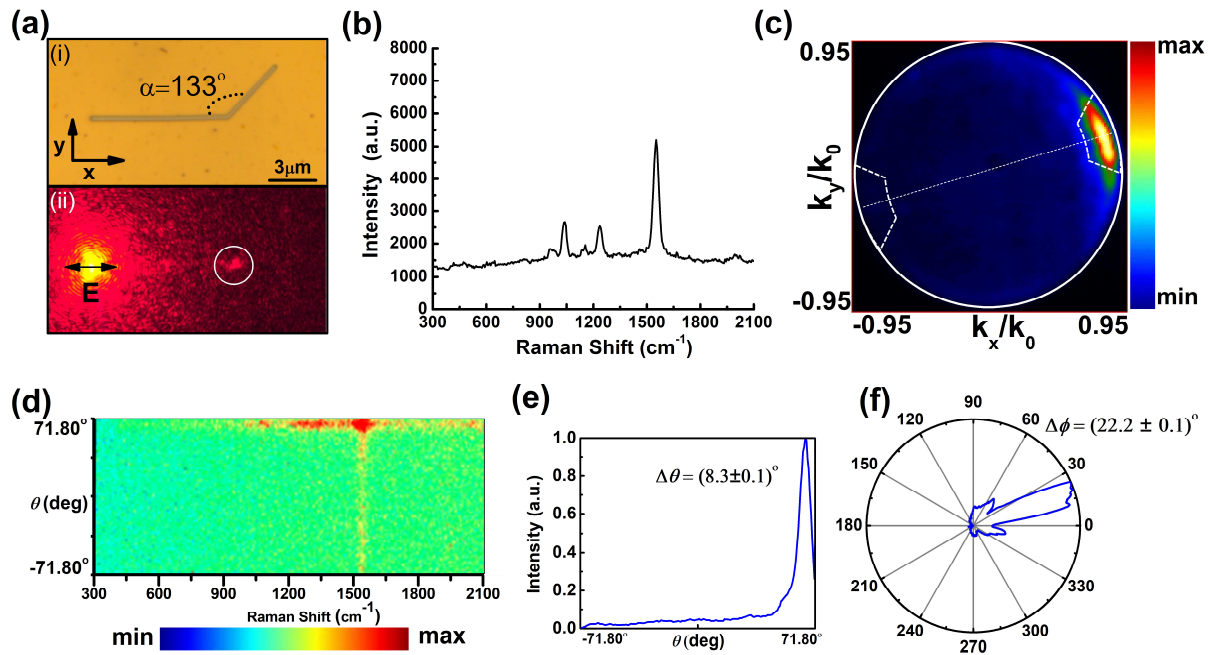

**Figure 5:** Directional SERS emission from B-NWoM cavity. (a) Bright field optical image of a bent-nanowire with an inter-arm angle α =133° placed on a gold mirror over which there is a self-assembled monolayer of BPT molecules. Elastic scattering image of the same bent-nanowire when one end was excited with 633 nm laser. (b) SERS spectrum from a monolayer of BPT molecules collected remotely from the white circle shown in (a)(ii). (c) Fourier plane image of SERS emission collected from the white region in (a)(ii) after rejecting elastic scattering light. (d) Energy-momentum imaging showing the directional out-coupling of the Raman lines and inelastic

background to higher angles. (e) Intensity cross-cut along the white dotted line in Fourier plane image showing the confinement of light with a FWHM of only 8.3°. (f) The intensity profile of azimuthal angles ($\phi$) for $\theta$ corresponding to maximum intensity in the Fourier plane image. The emission is very sharp with FWHM of only 22.2°.

Along with sharp SERS lines in the spectrum there is also an inelastic background emission from the PVP molecules of the nanowire,[38] which also gets captured undesirably in the Fourier plane image. To confirm that the emission at higher values was indeed SERS and not inelastic background emission, we performed energy-momentum imaging,[10,33] where selected portion of Fourier plane image is dispersed in the spectrometer to get the wavelength or energy information. We projected a part of Fourier plane image around the white dotted line to the spectrometer and dispersed it in terms of wavelength. The energy-momentum image (figure 5(d)) has greater intensity counts at higher k-values which confirms the out-coupling of both SERS and background from the kink at higher k-vectors. To quantify the emission, we used equation (1) to calculate the directionality of the emission. For the white dotted region in Figure 5(c), with $\delta_1$=7.5° and $\delta_2$=12.5°, the calculated directionality is (12.5 ± 0.1) dB which is greater than the maximum values reported for directionality calculation[30,39,40] for inelastic antennas because of the presence of single orientation of molecules in the B-NWoM cavity.

The intensity cross-cut (figure 5(e)) along the white dotted line in the Fourier plane image shows that the emission is very narrow in radial angles with a FWHM of only (8.3 ± 0.1)°. Furthermore, the intensity profile of azimuthal angles (figure 5f) ($\phi$) for $\theta$ corresponding to maximum intensity in the Fourier plane image shows that the emission has a FWHM of only (22.2 ± 0.1)° which is relatively less as compared to other reported values[30,39] because of the presence of only one orientation of molecules in the B-NWoM cavity. As compared to the emission wavevectors from B-NWoM using BPT molecular monolayer, the emission wavevectors from the dropcasted molecules are relatively broader when used either with nanowire on mirror or kink-nanowire on mirror cavities because of the preferred in-plane orientation of the dropcasted molecules (see supplementary information S10 and S11). The confinement of wavevector increases when silver nanowire is used with BPT molecular monolayer but because of finite reflection from the end of nanowire the back scattered light also increases[5] which reduces the directionality of emission (see supplementary information S12). Whereas, with B-NWoM geometry, the kink part of the bent-nanowire does not efficiently reflect the light in the backward direction and majority of the emission is only projected to the forward direction which results in a high unidirectionality of the emission. In addition, with B-NWoM geometry, the radiation pattern can also be tuned depending on the inter-arm angle. See supplementary information S13 for experiments and FDTD based simulations performed to study the variation of radiation pattern for elastic and SERS emission with a change in the inter-arm angle of the nanowire.

To conclude, we have shown beaming elastic and SERS emission with extremely narrow wavevector distribution using B-NWoM cavity. Instead of using conventional substrate with high refractive index, we utilized gold mirror for procuring beaming elastic scattering light from the kink of bent-nanowire. In addition, we utilized extended one-dimensional B-NWoM cavity formed between nanowire and mirror to design directional SERS antenna from a monolayer of vertically standing molecules sandwiched between nanowire and mirror. We used three-dimensional numerical simulations to corroborate the experimental results and to get insights on how to use the cavity for directing inelastic emission from molecules with a good control on the orientation of molecules. The B-NWoM geometry provides excellent angular confinement to the emission in the far-field with radial and azimuthal spreading with a FWHM of only 11.9° and 14.3° for elastic and 8.3° and 22.2° for SERS emission. The results presented here will readily be extrapolated to study and direct emission from two-dimensional materials placed in the elongated B-NWoM cavity. The strong electric field generated at the kink of the B-NWoM structures can also be used to remotely detect single molecule SERS signatures in fluid-phase and to study strong coupling physics.

**Acknowledgement**

Sunny Tiwari thanks Rohit Chikkaraddy (University of Cambridge, UK) for helping in the simulations performed and for fruitful discussions and Prof. Ramaprakash and Siddharth Maharana (IUCAA Pune, India) for fruitful discussion on imaging techniques. Authors also thank Chetna Taneja, Vandana Sharma, Utkarsh Khandelwal and Suryanarayan Banerjee (IISER Pune, India) for fruitful discussion. Sunny Tiwari also thanks Jean Paul Hugonin (Université Paris Saclay) for discussion on far-field calculations using reciprocity arguments.

This work was partially funded by Air Force Research Laboratory grant (FA2386-18-1-4118 R&D 18IOA118), DST Energy Science grant (SR/NM/TP-13/2016), and Swarnajayanti fellowship grant (DST/SJF/PSA02/2017-18) to G V PK.

**Supporting Information Available**

Experimental setup and sample preparation, details on FEM and FDTD simulations, experiment on Fourier plane imaging with dropcasted molecules in B-NWoM cavity, and experiments and simulations performed on the Fourier plane imaging of emission B-NWoM geometry with different inter-arm angles.

# Supplementary Information

# Beaming Elastic and SERS Emission from Bent-Plasmonic Nanowire on a Mirror cavity

*Sunny Tiwari*[1*], *Adarsh Bhaskar Vasista*[2], *Diptabrata Paul*[1], *Shailendra K. Chaubey*[1] *and G. V. Pavan Kumar*[1,#]

[1]Sunny Tiwari, Diptabrata Paul, Shailendra K. Chaubey and G. V. Pavan Kumar
Department of Physics, Indian Institute of Science Education and Research, Pune-411008, India

[2]Adarsh Bhaskar Vasista
Department of Physics and Astronomy, University of Exeter EX44QL, United Kingdom

[*]E-mail: sunny.tiwari@students.iiserpune.ac.in
[#]E-mail: pavan@iiserpune.ac.in

S1: Sample preparation and scanning electron microscope images of bent-nanowires with different inter-arm angles

S2: Experimental setup

S3: Fourier plane imaging of elastic emission from bent-nanowire on glass substrate

S4: Polarization resolved Fourier plane imaging of emission from the distal end of straight silver nanowire placed on a gold mirror

S5: Variation of emission wavevectors in Fourier plane images and change in $\Delta\phi$ with respect to analyzer angle

S6: Variation of analyzer angle for minimum $\Delta\phi$ spreading with a change in the inter-arm angle α of bent-nanowire

S7: Details on finite element method based calculations

S8: Fourier plane imaging of emission wavevectors from a chain of dipoles

S9: Details on finite difference time domain based calculations

S10: Fourier plane imaging of SERS emission collected remotely from the nanowire end with BPT molecules dropcasted on the mirror

S11: Fourier plane imaging of SERS emission from bent-nanowire on mirror cavity with BPT molecules dropcasted on the mirror

S12: Fourier plane imaging of SERS emission collected remotely from the nanowire end with a BPT molecular monolayer in the cavity

S13: Calculated Fourier plane imaging of elastic emission from bent-nanowire on mirror geometry and variation of $\phi_m$ for elastic and SERS emission with a change in the bending angle α of bent-nanowire

## S1: Sample preparation and scanning electron microscope images of bent-nanowires with different inter-arm angles

Sample preparation: Silver nanowires were synthesized using polyol process[1]. To bend the nanowire for making bent-NW, nanowire dispersed in ethanol solution were ultrasonicated for 30 seconds. The complete process of bending the nanowires using ultrasonication can be found in ref[2]. Figure S1 shows multiple bent-nanowire with different inter-arm angles. These bent-NWs were dropcasted on a 160 nm thick gold mirror for making B-NWoM geometry. For directing SERS emission from BPT molecules, a single layer of vertically oriented molecules was assembled on a 160 nm thick mirror using self-assembled monolayer technique. For preparing the self-assembled monolayer of BPT molecules on mirror, a gold coated glass cover slip was left in a solution of 1 mM solution of BPT molecules in ethanol for 24 hours. The cover slip was cleaned with ethanol before use for removing the molecules which were not assembled. Over the monolayer, B-NW was dropcasted for making B-NWoM cavity.

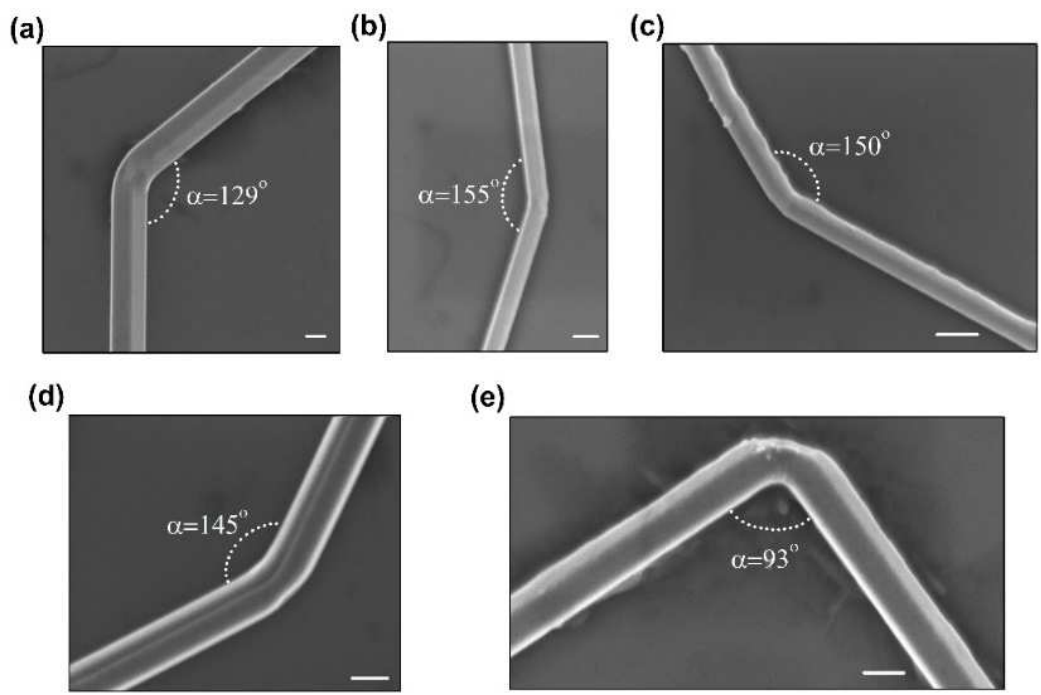

Figure S1: Scanning electron microscopy (SEM) image of bent silver nanowires (B-NW) with different inter-arm angles. (a-e) SEM image of B-NWs with bending angles, α=129°,155°,150°,145° and 93° respectively. Scale bar in (a-e) is 200 nm.

## S2: Experimental setup

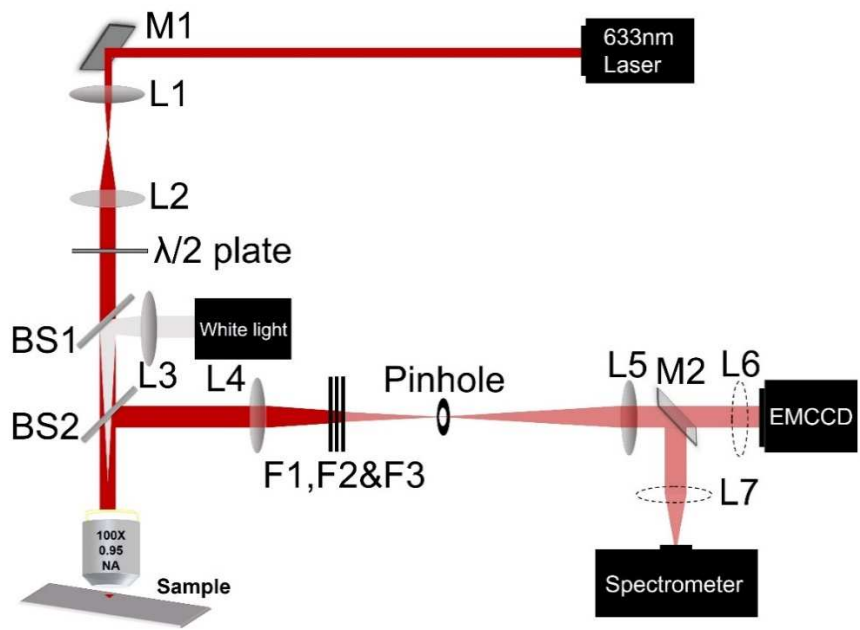

Figure S2: Schematic representation of the experimental setup.

The sample was excited using a high numerical aperture 100x, 0.95 NA objective lens. The backscattered light was collected using the same lens. The 633 nm laser light was expanded using a set of two lenses L1 and L2. M1 is a mirror. The polarization of the incoming laser was controlled by a λ/2 waveplate in the path. BS1 and BS2 are beam splitters to simultaneously excite the sample with laser and its visualization using white light. Lens L3 is used to loosely focus white light on the sample plane. F1, F2, and F3 are set of two edge filters and one notch filter to reject the elastically scattered light for SERS spectroscopy and Fourier plane[3-4] and energy-momentum imaging[3, 5]. Lenses L4 and L5 are used to project the emission to the Fourier plane onto the spectrometer or EMCCD. M2 is a flip mirror, used to project the light on the spectrometer for spectroscopy and energy-momentum imaging. Lenses L6 and L7 are flip lenses used to switch from real plane to Fourier plane.

**S3: Fourier plane imaging of elastic emission from bent-nanowire on glass substrate**

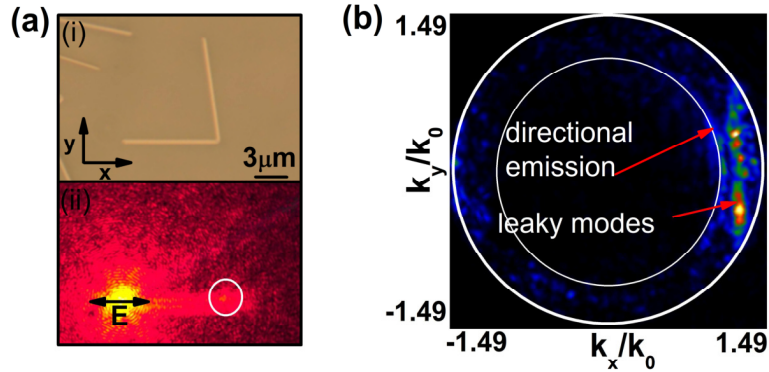

Figure S3: Fourier plane imaging of elastic emission from bent-nanowire placed on a glass substrate using a high numerical aperture oil-immersion objective lens. (a) (i) Bright field optical image of a bent-NW placed on a glass substrate. (ii) Elastic scattering image of the same bent-NW when one end of the nanowire is excited with 633 nm laser using 1.49 NA, 100x oil-immersion objective lens. (b) Fourier plane image of spatially filtered emission from the kink, captured using same objective lens. The directional emission and leaky modes of the nanowire are shown in the Fourier plane image.

Figure S3 shows the emission wavevectors of emission from the B-NW when the emission is collected through the glass side. (a) Optical image of a B-NW placed on a glass substrate. Upon excitation of one end of nanowire with a high numerical aperture objective lens, the nanowire plasmon polaritons out-couples from the kink part of nanowire as shown in the elastic scattering image of the same B-NW. The emission from the kink was spatially filtered and was projected to the EMCCD for Fourier plane imaging. The Fourier plane image shows two features which are directional emission from the kink part of the nanowire and the projection of the leaky modes of the nanowire which is a straight line above the glass-air critical angle, along $k_y/k_o$ at a constant $k_x/k_o$.

# S4: Polarization resolved Fourier plane imaging of emission from the distal end of straight silver nanowire placed on a gold mirror

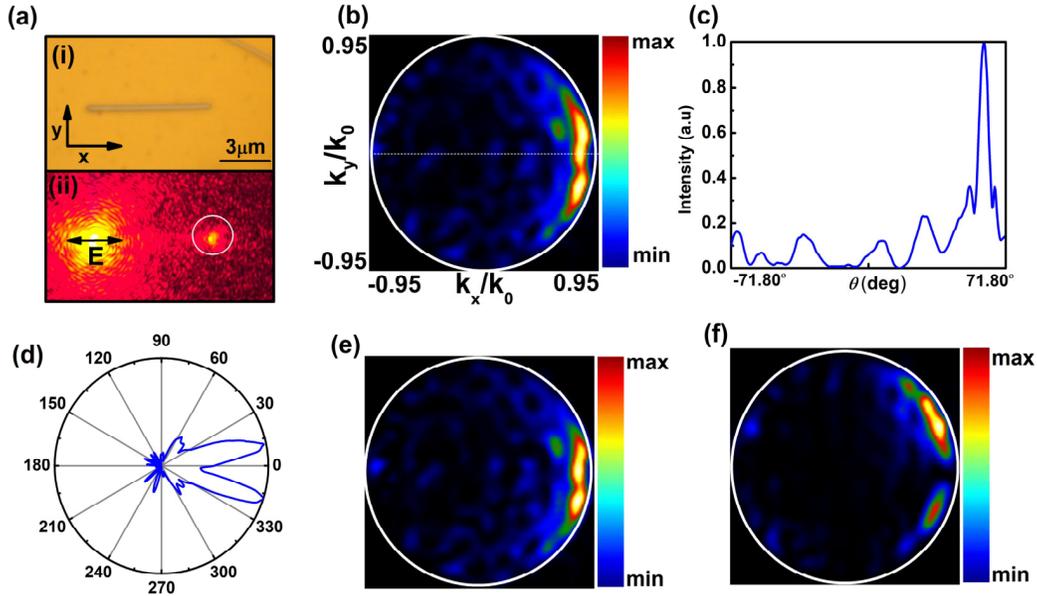

Figure S4: Polarization resolved Fourier plane imaging of elastic scattering emission from the distal end of straight silver nanowire placed on a gold mirror. (a) Optical imaging of a silver nanowire placed on a 160 nm thick gold mirror. (i) Bright field image of the nanowire. (ii) Elastic scattering image of the same nanowire when one end of the nanowire was excited with a 633 nm laser. The outcoupled light from the distal end was spatially filtered and was projected to the EMCCD for Fourier plane imaging. (b) Fourier plane image of the emission from the distal end. The emission is directed towards higher $+k_x/k_0$ values. (c) Intensity cross-cut along the $k_y/k_0=0$ line in the Fourier plane image shows the biasing of light in one direction in a narrow range of radial angles ($\theta$). (d) The intensity profile of azimuthal angles ($\phi$) for $\theta$ corresponding to maximum intensity in the Fourier plane image (b). (e) Polarization resolved Fourier plane image of emission from the distal end when the emission is analyzed along the length of nanowire. The $\phi$ spreading in the emission is narrower as compared to the unanalyzed emission. (f) Polarization resolved Fourier plane image of emission from distal end when the emission is analyzed transverse to the length of nanowire.

Figure S4 shows the polarization resolved Fourier plane imaging performed on the emission from the distal end of nanowire. The wavevector distribution of emission from the distal end is shown in the figure S4 (b). The emission is directed towards higher wavevectors. The intensity cross-cut along the white dotted line in the Fourier plane image shows that the emission is narrow in terms of radial angles. The intensity profile of azimuthal angles ($\phi$) for $\theta$ corresponding to maximum intensity in the Fourier plane image shows that the emission is broad in terms of $\phi$ spreading. When the emission is analyzed along and perpendicular to the nanowire axis, as shown in the polarization resolved Fourier plane image, (e) and (f), the wavevector distribution shows interesting behavior. The emission is more confined in terms of azimuthal angles when the emission is analyzed along the nanowire axis as compared to the analyzed emission perpendicular to the nanowire axis. The Fourier plane image shows that the emission is directed in two lobes and not one arc which is generally seen when the nanowire is placed on a glass substrate. Since, the substrate is gold and because of the large electric field in the cavity between the nanowire and mirror, the outcoupled light from the distal end of the nanowire is more intense transversely to the nanowire axis which is shown in our past work[6]. This make the emission to be directed in to two lobes in the Fourier plane image.

**S5: Variation of emission wavevectors in Fourier plane images and change in Δϕ with respect to analyzer angle**

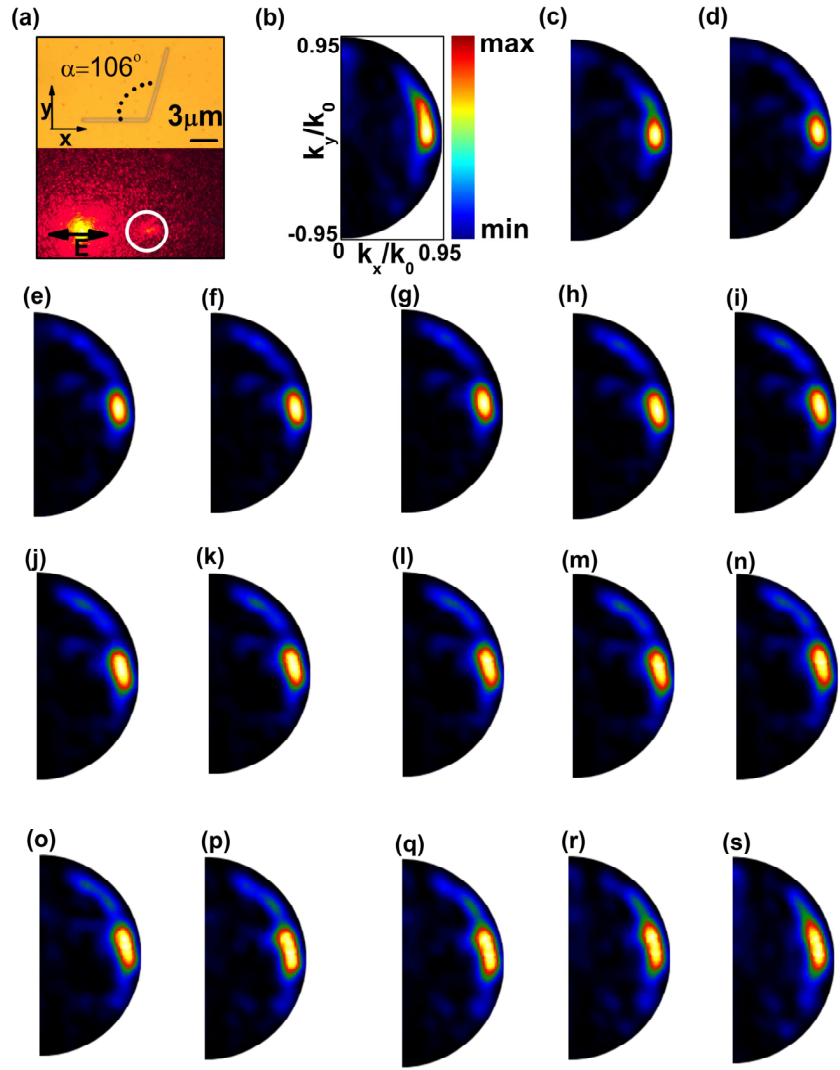

Figure S5: Variation of emission wavevectors in Fourier plane images with respect to analyzer angle. (a) (i) Bright field optical image of a bent-nanowire with an angle α=106° placed on a 160 nm thick gold mirror. (ii) Elastic scattering image of the same bent-nanowire when one end was excited with 633 nm laser. (b-s) Polarization resolved Fourier plane image of emission from kink part of the bent-nanowire when the emission was analyzed. The analyzer angle was changed from 35° to 205° with a step size of 10° with respect to the nanowire axis. The minimum ϕ spreading was obtained at an analyzer angle of 55° which is nearly half of α.

| Analyzer angle | $\Delta\phi$ (°) | Analyzer angle | $\Delta\phi$ (°) |
|---|---|---|---|
| No analyzer | 19.2 | 125 | 19.7 |
| 35 | 26.2 | 135 | 19.1 |
| 45 | 15.3 | 145 | 19.5 |
| 55 | 14.3 | 155 | 19.9 |
| 65 | 14.7 | 165 | 21.7 |
| 75 | 15.5 | 175 | 24.0 |
| 85 | 16.4 | 185 | 25.7 |
| 95 | 16.8 | 195 | 25.8 |
| 105 | 18.2 | 205 | 23.8 |
| 115 | 20.0 | | |

Table S5. Variation of $\Delta\phi$ with a change in the analyzer angle.

## S6: Variation of analyzer angle for minimum Δϕ spreading with a change in the inter-arm angle α of bent-nanowire

| S.No. | α (°) | Analyzer angle for minimum Δϕ spreading (°) |
|---|---|---|
| 1 | 59 | 40 |
| 2 | 106 | 51 |
| 3 | 113 | 80 |
| 4 | 124 | 50 |
| 5 | 125 | 65 |

Table S6: Variation of analyzer angle for minimum Δϕ spreading for bent-nanowires with different inter-arm angles α.

Table S6 shows the variation of analyzer angle for minimum Δϕ spreading for bent nanowires with different inter-arm angles α. For a bent-nanowire with acute bending angle 59°, the Δϕ spreading is minimum at 40°. For bending with obtuse angles, 106°,113°,124° and 125° the emission is roughly half the value of α. But even for slightly different inter-arm angles, the change in the value of Δϕ is large, which shows that the exact value depends on the geometry at the kink of the bent-nanowire and the complete description will need further investigation.

## S7: Details on finite element method based calculations

AgNW was modelled with a pentagonal cross-section with an edge to edge thickness of 350 nm and length of 10 μm. The gap between the nanowire and substrate was set to be 5 nm. This 5 nm gap is to model the PVP coating on the AgNW[7]. The refractive indices of the material were taken from ref[8]. The nanowire was excited with a focused Gaussian laser beam of wavelength 633 nm with polarization along the length of the nanowire. The plasmon propagation wavelength ($\lambda_{spp}$) for nanowire placed on glass substrate $\lambda_{spp}$ was calculated to be 557 nm, whereas for nanowire placed on the gold substrate $\lambda_{spp}$ was 613 nm.

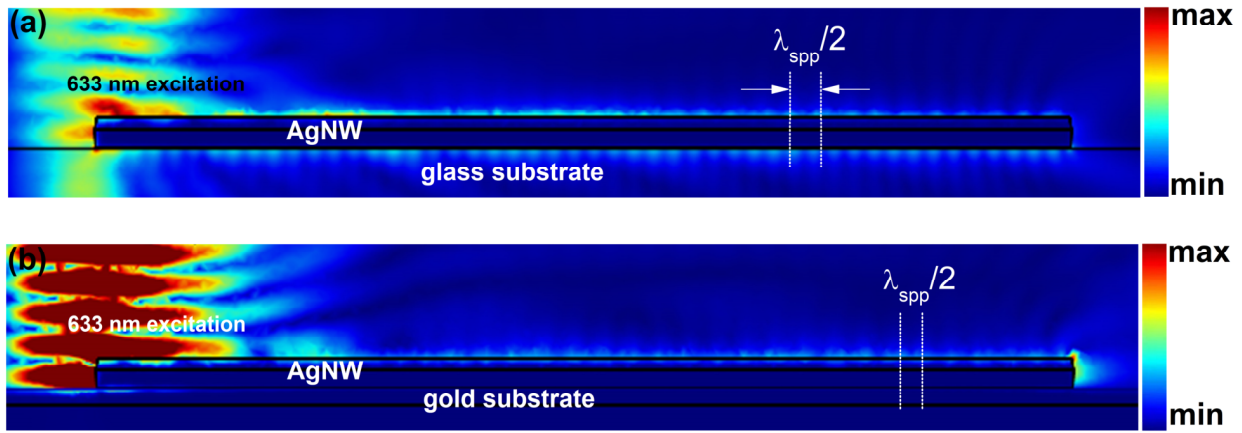

Figure S7: Calculated near-field electric field and plasmon propagation wavelength for nanowire placed on glass and gold substrate. Near-field electric field of nanowire placed on a glass (a) and gold (b) substrate. The nanowire was excited with a wavelength of 633 nm using a Gaussian excitation with polarization along the length of the nanowire. The plasmon propagation wavelength was 557 nm on a glass substrate and 613 nm on a gold substrate. The length of nanowire in (a) and (b) was 10 μm.

# S8: Fourier plane imaging of emission wavevectors from a chain of dipoles

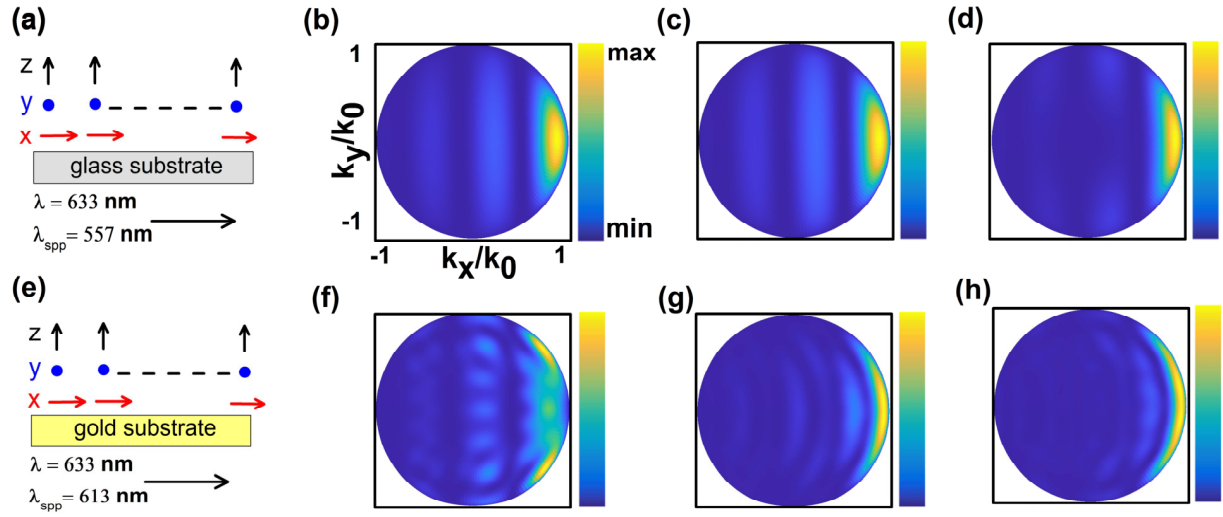

Figure S8: Fourier plane imaging of emission wavevectors from a chain of dipoles. A chain of 10 dipoles (x, y, or z oriented) placed at a distance of $\lambda/200$ from a glass (a) and gold (e) substrate. The phase retardation between the dipoles are set according to the plasmon propagation wavelength of 557 nm for glass substrate and 613 nm for gold substrate. Calculated Fourier plane images for a chain of x, y, or z oriented dipoles, placed on a glass substrate (b-d) and gold substrate (f-h) respectively. The complete length of the chain is 1 μm and the dipoles are oscillating at 633 nm.

Figure S8 shows the calculated Fourier plane images of emission wavevectors from a chain of dipoles. We choose a chain of 10 dipoles with a gap of 100 nm between two consecutive dipoles. The dipole moment is modulated by a phase factor of $\exp(ik_{spp}x_i)$ where $x_i$ is dipole position along the chain and $k_{spp}$ is $2\pi/l_{spp}$. Since, in the experiments we have used thick nanowires (~350 nm), we calculated plasmon propagation wavelength using COMSOL Multiphysics software (as discussed in S7). The dipoles are oscillating at a wavelength of $\lambda=633$ nm and are placed at a distance of $\lambda/200$ from the substrate. In each of the case, the emission is collected through the air side. The far-field patterns are calculated by projecting the calculated near-field electric field to the far-field using reciprocity argument[9].

On a glass substrate, the emission is more confined in radial angles when the orientation of the dipoles is along z directions as compared to the x and y oriented dipoles. This shows that the directionality and the confinement in the emission will be better where the effect of z dipoles is more. Also, for gold substrate, the far-field radiation pattern is confined in radial and azimuthal angles for z dipoles as compared to the x and y dipoles.

## S9: Details on finite difference time domain based calculations

Near-field electric field and far-field radiation pattern was simulated in Lumerical FDTD software. The nanowire was modelled as a cylindrical rod of length 6 um and radius of cylinder was set to be 125 nm. The end radius of the cylindrical rod was set to be 110 nm to match the geometry of the kink part of the bent-nanowire. The nanowire was placed on a 100 nm thick gold substrate and the gap between the nanowire and gold substrate was set to be 5 nm. This 5 nm gap is to model the PVP coating on the AgNW. The radiation pattern was calculated by projecting the near-field to the far field from a small region of the nanowire which contains the kink part of the bent-nanowire.

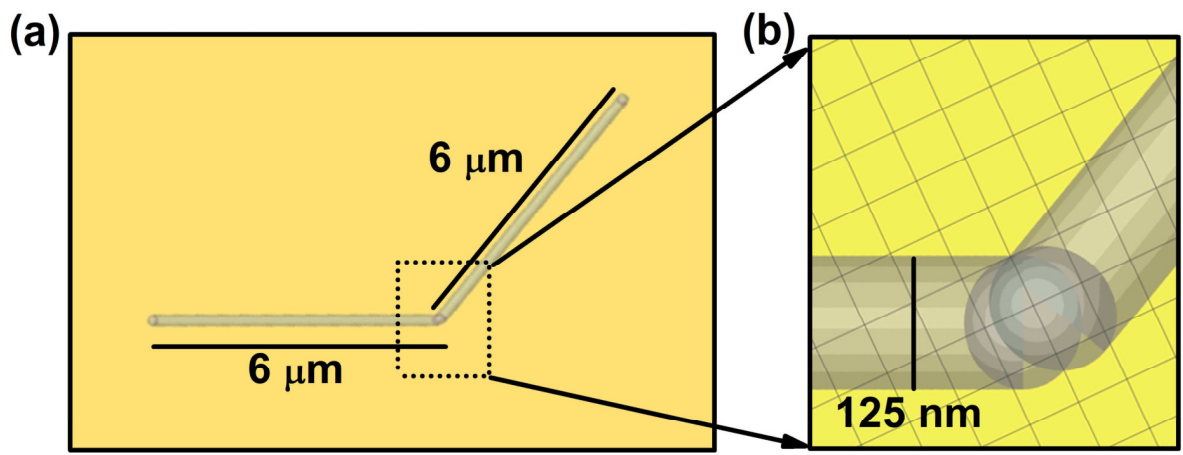

Figure S9: Geometry of the system used in FDTD simulations.

# S10: Fourier plane imaging of SERS emission collected remotely from the nanowire end with BPT molecules dropcasted on the mirror

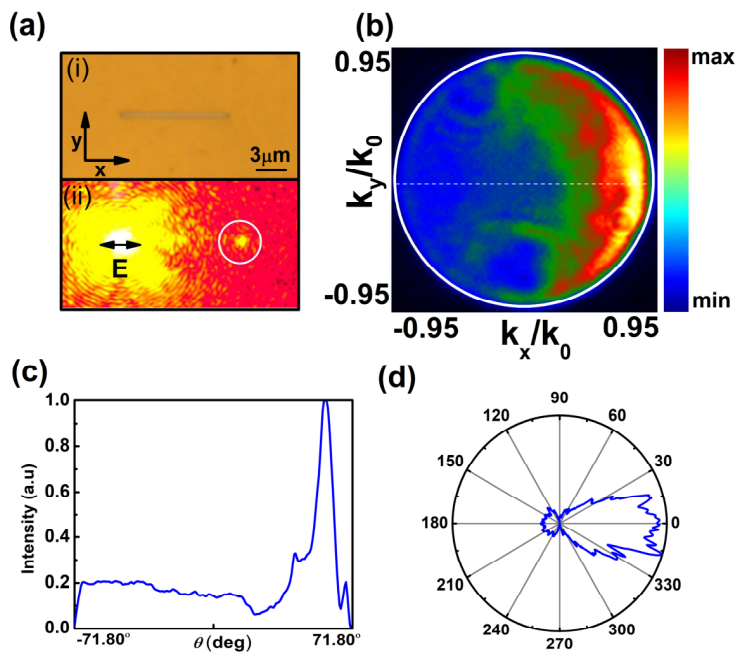

Figure S10: Fourier plane imaging of SERS emission collected remotely from the nanowire end with BPT molecules dropcasted on the mirror. (a) Optical image of a silver nanowire dropcasted on a 160 nm thick gold mirror. (i) Bright field optical image of nanowire placed on a gold mirror. (ii) Elastic scattering image of the same nanowire when one end of nanowire was excited using a 633 nm laser. (b) Fourier plane image of SERS emission collected remotely from the nanowire end using spatial filtering. (c) Intensity cross-cut along the $k_y/k_0=0$ line in the Fourier plane image shows the biasing of light in one direction in a narrow range of radial angles ($\theta$). (d) The intensity profile of azimuthal angles ($\phi$) for $\theta$ corresponding to maximum intensity in the Fourier plane image (b).

Figure S10 shows the Fourier plane imaging of SERS emission collected from the nanowire end. The BPT molecules are dropcasted on a 160 nm thick gold mirror and were left to dry. Silver nanowire was dropcasted on the BPT molecule coated gold mirror. One end of the nanowire was excited using a high numerical aperture objective lens and the out-coupled light from the distal end was collected using the same objective lens. The emission was projected was projected to the EMCCD for Fourier plane imaging. The Fourier plane image shows that the emission is directed towards higher wavevectors.

Since the molecules are dropcasted on the mirror, majority of the molecules are distributed in plane and the chances of molecules standing vertically on the mirror is negligible. Therefore, the emission is distributed broadly in terms of angles in the Fourier plane image. The intensity cross-cut (figure S10(c)) along the white dotted line shows that the emission is confined in terms of radial angles with a FWHM of 8.8°. But the backward emission is also prominent which reduces the directionality of the emission. The intensity profile of azimuthal angles (figure S10(d)) ($\phi$) for $\theta$ corresponding to maximum intensity in the Fourier plane image shows that the emission is very broad in azimuthal angles with a FWHM of 62.4°.

**S11: Fourier plane imaging of SERS emission from bent-nanowire on mirror cavity with BPT molecules dropcasted on the mirror**

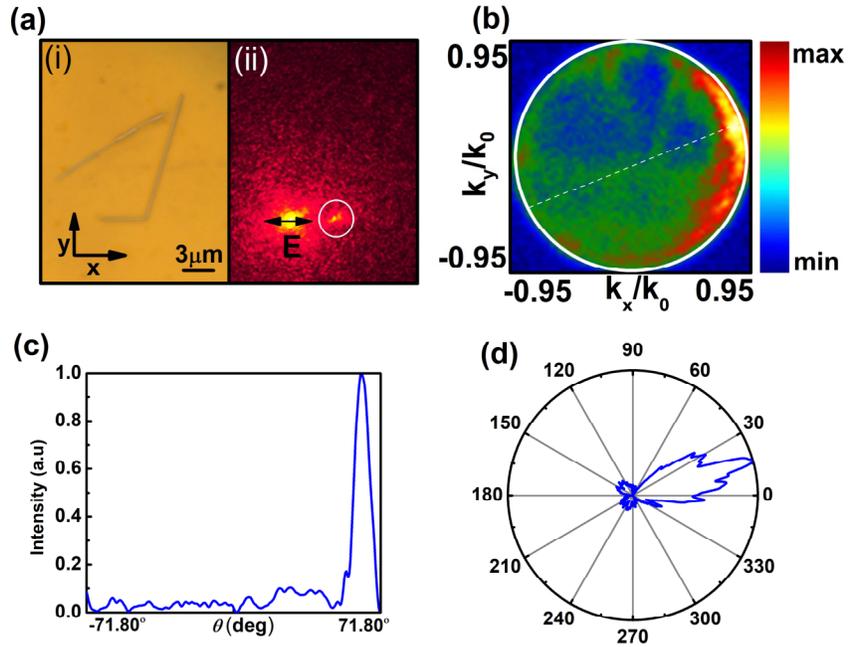

Figure S11: Fourier plane imaging of SERS emission from kink nanowire on mirror cavity with dropcasted BPT molecules. (a) Optical image of a bent-nanowire dropcasted on a 160 nm thick gold mirror. The inter-arm angle is 106°. The BPT molecules are dropcasted on the gold mirror. (i) Bright field optical image of bent-nanowire placed on a gold mirror. (ii) Elastic scattering image of the same bent-nanowire when one end of nanowire was excited using a 633 nm laser. (b) Fourier plane image of SERS emission collected remotely from the kink part of the bent-nanowire end using spatial filtering. (c) Intensity cross-cut along the $k_y/k_0=0$ line in the Fourier plane image shows the biasing of light in one direction in a narrow range of radial angles ($\theta$). (d) The intensity profile of azimuthal angles ($\phi$) for $\theta$ corresponding to maximum intensity in the Fourier plane image (b).

Figure S11 shows the Fourier plane imaging of SERS emission from dropcasted molecules on a 160 nm thick gold mirror in B-NWoM cavity. The preferred orientation of dropcasted molecules are in-plane and thus the directionality in the emission is because of x and y oriented dipoles which is relatively broad as compared to the z oriented dipoles (as shown in figure 5 of main manuscript). The cross-cut along the white dotted line in the Fourier plane image shows that the emission is confined in terms of radial angles (FWHM of 8.8°) but the emission in other regions is prominent. Similarly, the intensity profile of azimuthal angles ($\phi$) for $\theta$ corresponding to maximum intensity in the Fourier plane image (b) also shows relatively broad emission (FWHM of 42.2°) as compared to spreading with z oriented molecules as shown in figure 5 of main manuscript (FWHM of 22.2°).

**S12: Fourier plane imaging of SERS emission collected remotely from the nanowire end with a BPT molecular monolayer in the cavity**

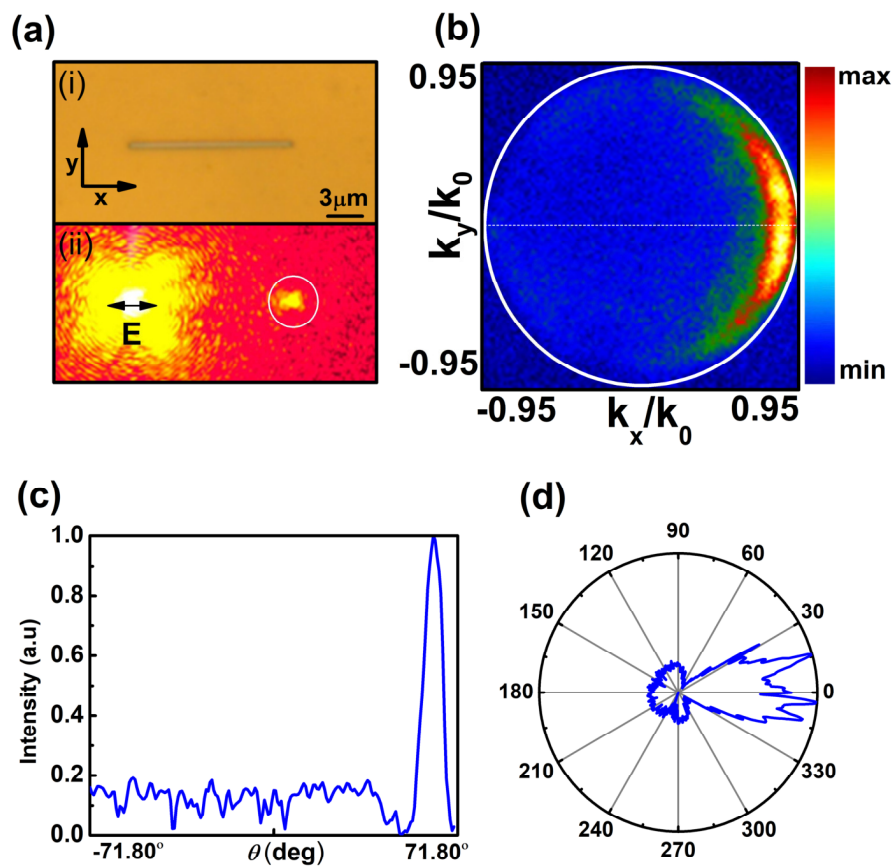

Figure S12: Fourier plane imaging of SERS emission collected remotely from the nanowire end. (a) Optical image of a silver nanowire dropcasted on a 160 nm thick gold mirror with a self-assembled monolayer of BPT molecules. (i) Bright field optical image of nanowire placed on a gold mirror. (ii) Elastic scattering image of the same nanowire when one end of nanowire was excited using a 633 nm laser. (b) Fourier plane image of SERS emission collected remotely from the nanowire end using spatial filtering. (c) Intensity cross-cut along the $k_y/k_0=0$ line in the Fourier plane image shows the biasing of light in one direction in a narrow range of radial angles ($\theta$). (d) The intensity profile of azimuthal angles ($\phi$) for $\theta$ corresponding to maximum intensity in the Fourier plane image (b).

Figure S12 shows the Fourier plane imaging of SERS emission collected from the nanowire end. Silver nanowire was dropcasted on a self-assembled monolayer of BPT molecule coated gold mirror. One end of the nanowire was excited using a high numerical aperture objective lens and the out-coupled light from the distal end was collected using the same objective lens. The emission was projected was projected to the EMCCD for Fourier plane imaging. The Fourier plane image shows that the emission is directed towards higher wavevectors.

Since the molecules are vertically orientated the emission is relatively narrow in terms of radial and azimuthal angles. The emission in the backward direction is still prominent in the backward direction. The intensity cross-cut along the white dotted line in the Fourier plane image shows that the emission is relatively narrow in radial angles (FWHM of 8.6°) as compared to the spreading when the molecules are dropcasted on the mirror. Even for azimuthal angles, the spreading reduces (FWHM of 49.4°) as compared to the case when the molecules are dropcasted.

**S13: Calculated Fourier plane imaging of elastic emission from bent-nanowire on mirror geometry and variation of $\phi_m$ for elastic and SERS emission with a change in the inter-arm angle α of bent-nanowire**

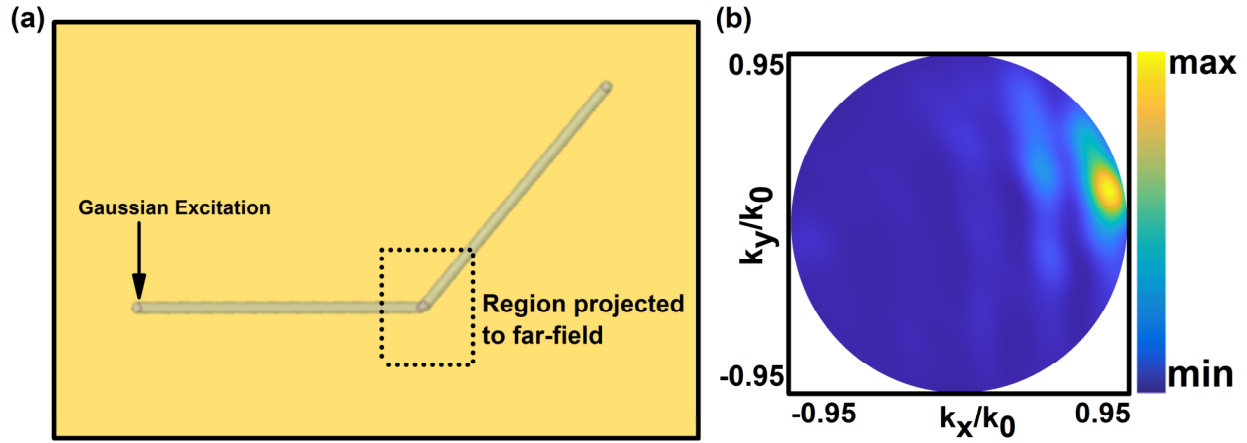

Figure S13: Calculated Fourier plane image of elastic emission from the bent-nanowire on mirror geometry. (a) Geometry used in the calculation as discussed in S9. The end of the nanowire is excited using a Gaussian excitation and the near-field from the kink part of bent-nanowire (shown in a rectangular dotted box) was projected to the far-field. (b) Calculated Fourier plane image of elastic emission from a bent-nanowire on mirror geometry with inter-arm angle 130° showing the out-coupled emission in a narrow range of wavevectors.

| S. No. | Elastic scattering (Experiment) α (°), $\phi_m$(°) | Elastic scattering (Simulation) α(°), $\phi_m$(°) | SERS emission (Experiment) α (°), $\phi_m$(°) | SERS emission (Simulation) α (°), $\phi_m$(°) |
|---|---|---|---|---|
| 1 | 84,180 | 84,174 | 50,178 | 50,187 |
| 2 | 93,179 | 93,174 | 69,179 | 69,172 |
| 3 | 106,171 | 106,176 | 108,157 | 108,175 |
| 4 | 114,169 | 114,172 | 133,156 | 133,165 |
| 5 | 119,174 | 119,173 | 135,154 | 135,167 |
| 6 | 125,152 | 125,174 | 152,158 | 152,170 |
| 7 | 143,168 | 143,173 | 156,166 | 156,168 |

Table S13: Variation of $\phi_m$ with a change in the inter-arm angle of bent-nanowire.

Table S13 shows the variation of $\phi_m$ (the azimuthal angle, where the emission is maximum in Fourier plane image) for elastic scattering and SERS emission (both experimental and FDTD calculations) with a change in the inter-arm angle of bent-nanowire. In the case of elastic scattering, for bent-nanowire with acute or right angle inter-arm angles, $\phi_m$ is along the length of nanowire. For obtuse angles, the emission changes from 152° (for α=125°) to 171° (for α=106°). The exact angular pattern depends on the geometry of the kink of the bent-nanowire. The calculations suggest that the emission is mainly directed towards angles slightly towards upper side of the nanowire and is always greater than 170°. For SERS, the emission is along the nanowire axis for acute angles, and for obtuse angles it varies from 157° (for α = 108°) to 166° (for α =156°). The calculations for radiation pattern for SERS emission shows a good agreement with the experimentally obtained data and the slight difference can be attributed because a slight change in the geometry can result in the radiation pattern at different angles.